\documentstyle[preprint,revtex]{aps}
\begin{document}
\draft
\preprint{bb: nucl-th/9502038, Nordita, Copenhagen, February 1995}

\begin{title}Correlations and Fluctuations in\\ High-Energy Nuclear
Collisions\end{title}
\author{Gordon Baym, B. Bl\"attel\thanks{Current address:
 Robert-Bosch GmbH, Abt.  ZWP, 7016 Gerlingen-Schillerh\"ohe, Germany}  }
\begin{instit}Department of Physics, University of Illinois at
     Urbana-Champaign\\ 1110 W. Green St., Urbana, IL 61801, USA
\end{instit}
\author{L. L. Frankfurt}
\begin{instit}Department of Physics, Tel Aviv University, Ramat Aviv, Israel
\end{instit}
\author{H. Heiselberg}
\begin{instit}NORDITA, Blegdamsvej 17, DK-2100 Copenhagen \O., Denmark
\end{instit}
\author{M. Strikman}
\begin{instit}Department of Physics, Pennsylvania State University,
      University Park, PA 16802, USA
\end{instit}

    \begin{abstract} \baselineskip=17pt Nucleon correlations in the target and
projectile nuclei are shown to reduce significantly the fluctuations in
multiple nucleon-nucleon collisions, total multiplicity and transverse energy
in relativistic heavy-ion collisions, in particular for heavy projectile and
target.  The interplay between cross-section fluctuations, from color
transparency and opacity, and nuclear correlations is calculated and found to
be able to account for large fluctuations in transverse energy spectra.
Numerical implementation of correlations and cross-section fluctuations in
Monte-Carlo codes is discussed.
\end{abstract}

\baselineskip=18pt
\section{Introduction}

    Ultrarelativistic heavy-ions experiments at CERN and Brookhaven have
provided important information on the processes involved in high-energy
nucleus-nucleus collisions.  Global features such as stopping, multiplicity,
transverse energy, and rapidity distributions of particles can be described to
a good approximation by models based on multiple hadronic collisions.
However, several results in central nucleus-nucleus collisions, such as
strangeness enhancement \cite{WA85,NA35}, $J/\Psi$ enhancement \cite{NA38},
and large fluctuations in multiplicity and transverse-energy distributions
\cite{NA34,BFS}, deviate from naive extrapolations based on summing
nucleon-nucleon or nucleon-nucleus collisions.  The crucial issue is to
determine whether these interesting signals arise from the presence of a
quark-gluon plasma or from more conventional phenomena such as rescattering
and thermalization processes in the hot hadronic gas formed in the wake of the
collisions.  As we describe here, study of fluctuations provides a
particularly valuable probe of the physics underlying these processes.

    An early indication of the importance of understanding fluctuations in
observables was provided by the NA34 geometrical model \cite{NA34}.  This
model gave reasonably accurate parametrizations of the NA34 multiplicity and
transverse energy ($E_T$) spectra in terms of summing multiple nucleon-nucleon
collisions -- each of which provides a distribution of transverse energy and
multiplicity -- with the collision geometry incorporated.  However, as Baym et
al.  \cite{BFS} pointed out, the fluctuations extracted in this way from
experiment, as reflected in the widths of the high-energy tails of $E_T$
distributions, are much larger than those expected from a microscopic model
with the known fluctuations in transverse energy in nucleon-nucleon
collisions.

    We focus here on understanding the subtle and interlocking roles of the
nucleon correlations in the underlying nuclear structure of the colliding
nuclei, the range of the high energy nucleon-nucleon interactions, and
cross-section fluctuations in determining fluctuations in observables.  In
Refs.  \cite{PRL91} and \cite{QM91} we gave a brief account of the importance
of including cross-section fluctuations and nuclear correlations in
calculating $E_T$ spectra.  Calculations with these effects included showed
that one should expect the large fluctuations apparent in the data.  The
present work includes a more detailed derivation of these effects.

    Short-range correlations, which arise mainly from the short-range
repulsion between nucleons as well as Pauli blocking, exclude other nucleons
within a distance of order the correlation length, and reduce the fluctuations
around the mean number, $\langle N\rangle$, of nucleon-nucleon (NN) collisions
that a projectile makes while traversing the target nucleus.  For example, a
high energy proton making a central collision with a heavy nucleus, e.g.,
$^{208}Pb$, interacts on average with $\langle N\rangle=2\sigma\rho_0 R\simeq
7$ target nucleons.  The interaction zone is small compared to the volume of
the nucleus but large compared to that of a single nucleon.  For a target
nucleus containing an uncorrelated gas of nucleons freely running about, the
number of target nucleons in the interaction zone can range from none to all,
as given by the binomial distribution, leading to large flucutations in $N$.
However, the nuclear matter in the nucleus is in fact a strongly correlated
self-bound liquid, not a gas; the density fluctuations are shorter ranged, and
the fluctuations in the number of nucleons in the interaction zone greatly
reduced.  This effect is even more significant for nucleus-nucleus collisions.

    Since the correlation length is of order the internucleon distance, itself
of order the range of the interactions or the cross-sectional distance
$r_s\equiv (\sigma/\pi)^{1/2}$, fluctuations in observables depend as well on
the detailed interplay of the correlations and the range of the interactions.
A normal assumption in microscopic models is that the projectile interacts
with all nucleons in the tube along its trajectory with area given by the
total NN cross section.  However, allowing a more diffuse interaction range
tends to reduce the importance of correlations, and increase the fluctuations,
as will become evident below.

    We primarily develop the study of correlations and fluctuations within the
framework of binary collisions of target and projectile nucleons, but we
briefly look as well at the predictions of the ``wounded-nucleon" description
of sources in terms of participant nucleons.  The underlying basis of the
binary-collision picture is the cutting rules in the field-theoretic
diagrammatic analysis of high-energy scattering processes that were found by
Abramovski{\u i}, Gribov, and Kancheli (AGK) \cite{AGK} and Bertocchi and
Treleani \cite{BT} within the inelastic eikonal approximation.  These rules
imply exact cancellations between inelastic interactions of the projectile
with more than one nucleon at a time.  They thus imply that in nucleon-nucleus
scattering, away from the projectile fragmentation region, inclusive particle
production is that in NN scattering times the number of nucleons, $A$, in the
target nucleus; in nucleus-nucleus (BA) scattering away from the nuclear
fragmentation regions, the result is that in NN scattering times $BA$, the
numbers of nucleons in the projectile and target.  In the fragmentation
regions, shadowing effects must also be taken into account.  In the Appendix
we discuss the kinematical regions where the AGK theorem is valid; we also
indicate how one can establish the correspondence between the basic
probabilistic formulae deduced in this paper from geometrical arguments and
the field-theoretic combinatorics of the AGK
approach.\footnote{\baselineskip=15pt The experimental situation does not make
a clear case for either an underlying binary-collision or wounded-nucleon
description.  In addition to the elementary collision processes described by
the AGK theorem, experiments are sensitive to secondary interactions
(rescattering) as well as coherent processes, e.g., showers of particles
overlapping spatially and interacting strongly, which the binary picture does
not account for.  Proportionality of the inclusive spectrum to $A$ at
mid-rapidities is observed experimentally, e.g., in the recent systematic
study by Whitmore et al. of pA scattering at 100 and 320 GeV lab energy
\cite{Whitmore}.  The faster than $\sim A$ dependence for low total momentum
secondary hadrons observed near the nuclear fragmentation regime is likely due
to secondary hadron interactions, including interaction of produced hadrons
with residual nucleons.  On the other hand, experimental $^{16}$O+Em(ulsion)
data \cite{Evert} indicate that total multiplicity is intermediate that of the
binary-collision model and the wounded-nucleon model.}

    Cross-section fluctuations also increase the fluctuations in the number of
collisions, as shown in \cite{PRL91}.  The basic picture is that, owing to
frozen color configurations in the projectile, the probability of a given
cross section, $\sigma$, in a primary collision is given by a distribution
function, $P(\sigma)$, instead of the cross section always taking a constant
(mean) value.  In Ref.  \cite{hadr} we discussed in detail the nature of
cross-section fluctuations in nucleon-nucleon interactions, and for
pion-nucleon interactions in Ref.  \cite{BBFS93}, and showed how the
distribution function $P$ for cross section fluctuations can be determined
from both inelastic shadowing in nucleon-deuteron scattering, and from
diffractive excitation in high-energy nucleon-nucleon scattering.  In the
Appendix we briefly discuss cross-section fluctuations within the AGK
framework.

    This paper is organized as follows.  In Sec. 2 we first calculate the
effect of nuclear correlations on fluctuations in the number of multiple NN
interactions in proton-nucleus collisions.  We then generalize the results to
nucleus-nucleus collisions in Sec. 3. In the following section, 4, we extend
the analysis to include contributions from cross-section fluctuations, and
then, in Sec. 5 we apply these results to the question of the fluctuations in
the tails of multiplicity and transverse energy distributions.  In the final
section we summarize and outline the procedure necessary to implement the
effects discussed here in the Monte-Carlo descriptions of AA collisions, and
in the Appendix discuss the AGK theorem.

\section{Correlations and fluctuations in proton-nucleus collisions}

    To gain a first understanding of the effect of nuclear correlations on
fluctuations we begin by examining collisions of protons with heavy nuclei,
and treat the more complicated case of nucleus-nucleus collisions below.
Neglecting cross-section fluctuations at this point, we take the NN inelastic
cross-section to be its average value, $\sigma=\bar{\sigma}\simeq 32$mb (at
CERN energies).

    The mean number of nucleon-nucleon collisions, $N$, that a proton at
impact parameter ${\bf b}$ makes with the target nucleus can be written, in
the impact parameter formalism of Ref.  \cite{MP}, as
\begin{eqnarray}
    \langle N\rangle = \int d^3 r \rho_A(r)
  \frac{d\sigma}{d^2{b}}({\bf b}-{\bf r}_\perp),
\label{N}
\end{eqnarray}
where $\rho_A$ is the nuclear density distribution, ${\bf r}_\perp$ is the
transverse coordinate, and ($d\sigma/d^2b)({\bf b}-{\bf r}_\perp)$ is the
probability for the projectile nucleon with impact parameter ${\bf b}$ to
collide with a target nucleon at position ${\bf r}$.  In the eikonal
approximation\footnote{
 \baselineskip=15pt
    By eikonal here we mean that projectile nucleons are only allowed to
scatter on target nucleons and visa versa.}
with straight-line geometry for multiple NN collisions -- the Glauber
approximation -- $d\sigma/d^2b$ depends only on the transverse coordinates
${\bf r}_\perp$; it has an interaction range $\sim r_s\simeq$ 1 fm and obeys
the sum rule
\begin{eqnarray}
    \int d^2 r_\perp \frac{d\sigma}{d^2{b}}({\bf r}_\perp) = \sigma.
\label{Psum}
\end{eqnarray}
As we shall see below, it is important to take the range of the
interaction into account in the multiple NN scattering problem since it is
comparable to the average internucleon distance
$r_0=(3/4\pi\rho_0)^{1/3}\simeq 1.16$ fm (where $\rho_0\simeq$ 0.16 fm$^{-3}$
is nuclear matter density), as well as the correlation length, $\xi$, between
nucleons in a nucleus, $\simeq$ 1 fm.  Since $r_s\ll R_A$, the target radius,
the mean number of NN interactions can, aside from a small correction at the
nuclear surface, be calculated without detailed knowledge of $d\sigma/d^2b\,$;
using (\ref{Psum}) we have
\begin{eqnarray}
    \langle N\rangle= \sigma\int_{-\infty}^{\infty} dz
    \rho_A({\bf b},z) \equiv \sigma T_A({\bf b}),
    \label{N1}
\end{eqnarray}
where $T_A({\bf b})$ is the thickness function for the target nucleus with
mass number $A$.  If the nucleus is assumed to have constant density $\rho_0$
with radius $R_A=r_0A^{1/3}$, then $T_A(b)=2\rho_0\left(R_A^2
-b^2\right)^{1/2}$.

    The fluctuation in the number of collisions, $N$, at impact parameter
${\bf b}$, is similarly given by\footnote{
\baselineskip=15pt
    In writing (\ref{N2a}) we neglect Glauber shadowing in the interaction of
the projectile with several nucleons of the nucleus.  As discussed in the
Appendix, AGK cancellation \cite{AGK,BT} in the Glauber model justifies
this neglect when calculating average characteristics of the final
state away from the projectile fragmentation region.}
\begin{eqnarray}
     \langle N(N-1)\rangle &=& \int d^3 r d^3 r'\rho^{(2)}_A({\bf r},{\bf r}')
    \frac{d\sigma}{d^2{b}}({\bf b}-{\bf r}_\perp)\frac{d\sigma}{d^2{b}}
    ({\bf b}-{\bf r}_\perp ');
    \label{N2a}
\end{eqnarray}
here $\rho_A^{(2)}$ is the two-body density distribution
\begin{eqnarray}
\rho_A^{(2)}({\bf r},{\bf r'})
   &=& \langle \psi^\dagger({\bf r}) \psi^\dagger({\bf r}')\psi({\bf r}')
          \psi({\bf r})\rangle  \nonumber\\
   &\equiv& \rho_A(r)\rho_A(r') \left(1-C_A({\bf r},{\bf r}')\right);
\label{rho2}
\end{eqnarray}
where $\psi^\dagger$ and $\psi$ are the nucleon creation and annihilation
operators.  Because nucleons tend to stay away from each other at short
distances, each nucleon is surrounded by a {\it correlation hole}, described
by the correlation function $C_A({\bf r},{\bf r}')$, which generally, as a
function of ${\bf r'}$, is non-vanishing about the nucleon at ${\bf r}$ over a
range $\xi$ in $|{\bf r}-{\bf r}'|$ of order 1 fm in nuclear matter.  From the
commutation relations of $\psi$ and $\psi^\dagger$ one readily derives the sum
rule on the correlation hole
\begin{eqnarray}
    \int d^3r'\rho_A(r')C_A({\bf r},{\bf r}')=1,
\label{C}
\end{eqnarray}
which says that the hole depletes precisely one nucleon.  Equation
(\ref{rho2}) implies that $C_A\le 1$.  For a nuclear liquid, one
expects
$C_A$ to be non-negative, as is borne out by detailed calculations
\cite{Vijay}.\footnote{\baselineskip=15pt
    One can imagine situations in which $C_A$ could be negative, e.g., halo
nuclei ($^{11}$Li) or helium clustering inside light nuclei.  In these systems
high density ``lumps" inside the nucleus increase the density fluctuations
tending to make $C_A$ negative at small separations.  However, such effects
are overwhelmed by short range correlations.}

    With (\ref{rho2}), we can write (\ref{N2a}) as
\begin{eqnarray}
  \langle N(N-1)\rangle &=& \int d^3rd^3r'\rho_A(r)\rho_A(r')
                  (1-C_A({\bf r},{\bf r}'))
    \frac{d\sigma}{d^2{b}}({\bf b}-{\bf r}_\perp)\frac{d\sigma}{d^2{b}}
     ({\bf b}-{\bf r}_\perp ')
      \nonumber\\
            &\equiv&
        \left( \langle N\rangle^2-\langle N\rangle\alpha \right),  \label{N2}
\end{eqnarray}
where we define the dimensionless parameter $\alpha$ by
\begin{eqnarray}
   \alpha =  \langle N\rangle^{-1}
        \int d^3{\bf r}\: d^3{\bf r}'\rho_A(r)\rho_A(r')
        C_A({\bf r};{\bf r'})
        \frac{d\sigma}{d^2{b}}({\bf b}-{\bf r}_\perp)\frac{d\sigma}{d^2{b}}
    ({\bf b}-{\bf r}_\perp ').
      \label{alfa}
\end{eqnarray}
Physically, the parameter $\alpha$ measures the fraction of the overlap
between the collision volume and the volume of the correlation hole described
by $C_A$ (Fig. 1).  To see this let us take the correlation function to be
spherically symmetric, i.e., $C_A({\bf r},{\bf r}')=C_A(|{\bf r}-{\bf r}'|)$.
Noting that for large nuclei both $r_s$ and $\xi$ are much smaller than the
nuclear radius, and for non-peripheral collisions the integral in (\ref{alfa})
is dominated by the bulk of the nucleus and is independent of impact
parameter, we see that Eq.  (\ref{alfa}) reduces to
\begin{eqnarray}
    \alpha = \frac{\rho_0}{\sigma} \int dz d^2 r_\perp d^2 r'_\perp
             C_A\left(\sqrt{z^2+({\bf r}_\perp-{\bf r}'_\perp)^2}\right)
             \frac{d\sigma}{d^2{b}}({\bf r}_\perp)
             \frac{d\sigma}{d^2{b}}({\bf r}'_\perp) \, . \label{alfap}
\end{eqnarray}

    From Eq. (\ref{N2}) we obtain the variance in the number of
collisions,
\begin{eqnarray}
  \omega_N\equiv \frac{\langle N^2\rangle-\langle N\rangle^2}{\langle N\rangle}
          =  1-\alpha \, .\label{o2}
\end{eqnarray}
This equation shows that nuclear correlations, through the quantity
$\alpha$, reduce the fluctuations in proton-nucleus collisions; we shall see
below that nucleus-nucleus collisions are even more sensitive to nuclear
correlations.  The parameter $\alpha$ varies from zero, for an uncorrelated
gas of nucleons, to unity, for a strongly correlated system, independent of
the detailed forms of the correlation and collision probability functions.
It is instructive first to evaluate $\alpha$ in various limits, and then
consider the general case.

\subsection{Extreme optical limit}

    In the extreme optical limit in which the range of the interation $r_s$ is
negligible, i.e., $r_s\ll \xi$, the NN scattering probability, $d\sigma/d^2b$,
can be approximated by
\begin{eqnarray}
    \frac{d\sigma}{d^2{b}}({\bf b}-{\bf r}_\perp) \simeq
    \sigma \delta^{(2)}({\bf b}-{\bf r}_\perp) \, .
    \label{P}
\end{eqnarray}
Then from Eq. (\ref{alfap}) we find that
\begin{eqnarray}
   \alpha  \simeq \sigma\rho_0 \int dz C_A(z) \, ,\quad
   r_s\ll\xi.
\label{depth}
\end{eqnarray}
The parameter $\alpha$ can thus be interpreted as the {\it optical depth}
of the correlation hole, the width in nucleon mean-free paths,
$1/\sigma\rho_A$, across the hole.

    If the correlation function is a simple step function, $C_A(|{\bf
r-r'}|)=\Theta(\xi-|{\bf r-r'}|)$, then the sum rule (\ref{C}) implies that
$\xi=r_0=(3/4\pi\rho_0)^{1/3}$, and we obtain
\begin{eqnarray}
   \alpha\simeq \frac{3}{2}\frac{\sigma}{\pi \xi^2} \, ,
   \quad r_s\ll\xi.
    \label{axi}
\end{eqnarray}

\subsection{Ideal gas}

    In an ideal gas with $A$ distinguishable particles, the correlation
function is a constant, $C_A=1/A$, determined by the sum rule; the correlation
``hole" extends without structure across the entire nucleus (see Fig. 2).  In
this case we can also use Eq.  (\ref{P}) and find from Eq.  (\ref{alfa}) that
for collisions at small impact parameter,
\begin{eqnarray}
    \alpha=\langle N\rangle/A \, ,     \label{agas}
\end{eqnarray}
which coincides with the result (\ref{axi}) for $\xi=R$.  The term
1-$\alpha$ in Eq.  (\ref{o2}) is then the standard binomial result for the
fluctuations in the number of collisions of a projectile passing through an
uncorrelated nuclear gas, with individual collision probability $p=\langle
N\rangle/A=\alpha$.  This is the minimum value for $\alpha$, corresponding to
maximum fluctuations, and correlation length $\xi$ of order the nuclear
radius.

\subsection{Maximum correlations}

    When $\xi \ll r_s$ we can write
\begin{eqnarray}
   \rho_A(r)C_A({\bf r,r'})\simeq\delta^{(3)}({\bf r-r'})
       \, , \label{maxcor}
\end{eqnarray}
and find from Eq. (\ref{alfa}) that
\begin{eqnarray}
 \alpha =  \frac{1}{\sigma} \int d^2b \left(\frac{d\sigma}{d^2{b}}({\bf b})
     \right)^2
           \, ,  \quad \xi\ll r_s\, . \label{as}
\end{eqnarray}
If we assume, as one often does in event generators simulating multiple NN
collisions, that the projectile nucleon collides with all nucleons situated
within a tube of cross-sectional area $\sigma$ along the beam axis with impact
parameter ${\bf b}$ (see Fig. 1), i.e.,
\begin{eqnarray}
    \frac{d\sigma}{d^2{b}}({\bf b}-{\bf r}_\perp) =
    \Theta(r_s-|{\bf b}-{\bf r}_\perp|) \, ,\label{st}
\end{eqnarray}
we find
\begin{eqnarray}
 \alpha = 1 \, ,  \quad \xi\ll r_s, \label{amax}
\end{eqnarray}
the maximum value of $\alpha$.  As Eq.  (\ref{o2}) shows, the fluctuations
in the number of binary collisions vanishes in this case.  This result is due
to the fact that in this limit each nucleon is confined to its correlation
hole, and for $\xi\ll r_s$ the number of nucleons within the tube is
determined by the ratio of the volume of the tube to the volume of the
correlation hole and cannot fluctuate around that value.

\subsection{Gas with excluded volume}

    In event generators, nucleon correlations are as a rule approximated by
not allowing two nucleons to be situated within a distance, $\tilde{\xi}$, of
each other.  The correlation function becomes in this way that of a gas with
an excluded volume (see Fig. 2), which to a first approximation, when
$\tilde{\xi}\le r_0$, is given by:  $C_A=1$ for $r\le\tilde{\xi}$ and, from
the sum rule, $C_A=(r_0^3-\tilde{\xi}^3)/R_A^3$ for $r>\tilde{\xi}$.  Since
$\alpha$ is linearly proportional to the correlation function, it is the
weighted sum of the two pieces.  To evaluate the first term, we use Eq.
(\ref{amax}) when $\tilde{\xi}\ll r_s$, and Eq.  (\ref{depth}) when
$\tilde{\xi}\gg r_s$; for the second term we use Eq. (\ref{agas}) generally,
and find in the two extreme limits,
\begin{eqnarray}
    \alpha =\left\{ \begin{array}{ll}
    &\frac{\tilde{\xi}^3}{r_0^3}
       + \left(1-\frac{\tilde{\xi}^3}{r_0^3}\right) \frac{\langle N\rangle}{A}
       \, , \quad \tilde{\xi}\ll r_s \,  \nonumber\\
           & \frac{3}{2}\frac{\sigma\tilde\xi}{\pi r_0^3}
       + \left(1-\frac{\tilde{\xi}^3}{r_0^3}\right) \frac{\langle N\rangle}{A}
       \, , \quad \tilde{\xi}\gg r_s \, .
       \end{array}\right.
\label{aexcl}
\end{eqnarray}
A numerical calculation for general $\tilde{\xi}/r_s$ with the ``tube"
cross-sectional area of Eq.  (\ref{st}) results in fluctuations
well-approximated by the first line of Eq.  (\ref{aexcl}), except for larger
values of $\tilde{\xi}$, where $\alpha$ is reduced by approximately 20\%.  It
is important to note that the fluctuations and $\alpha$ are very sensitive to
the choice of $\tilde{\xi}$.  In the following subsection we treat more
generally the situation where the range of the correlation function is
comparable to $r_s$.

\subsection{General case}

    The nucleus is a saturated self-bound liquid and is thus strongly
correlated, as can be seen from the correlation function of Fig. 2 taken from
Ref.  \cite{Vijay}.  Since the correlation length $\xi$ is of order the
internucleon distance $r_0=(3/4\pi\rho_0)^{1/3}\simeq 1.17$ fm, itself of
order $r_s\simeq 1.0$ fm, $\alpha$ depends on the detailed structure of $C_A$
and $d\sigma/d^2b$.  To make the discussion of the general situation more
specific, we construct simple parametrizations of both the correlation
function and the scattering probability, $d\sigma/d^2b$.  In particular we
assume a Gaussian form for the correlation function (see Fig. 2),
\begin{eqnarray}
    C_A({\bf r,r}') = e^{\displaystyle{-({\bf r-r}')^2/2\xi^2}} \, ;
    \label{Cg}
\end{eqnarray}
the sum rule (\ref{C}) relates the correlation length $\xi$ to the nuclear
density by
\begin{eqnarray}
    \xi =  r_0 (2/9\pi)^{1/6} \simeq 0.75\, {\rm fm} \, . \label{xir0}
\end{eqnarray}
[The results here can easily be generalized to a correlation function
consisting a sum of several Gaussians, which more closely fits calculated
correlation functions, e.g., in Ref. \cite{Vijay}.  See Fig. 2.]

    In the impact parameter formalism of Ref.  \cite{MP}, the imaginary part
of the scattering amplitude, Im $f(q)$, for momentum transfer $q$, is related
to the collision probability function by
\begin{eqnarray}
{\rm Im} f(q) = \frac{k}{4\pi}\int d^2 b e^{i{\bf q}\cdot{\bf b}}
 \frac{d\sigma}{d^2{b}}({\bf b}) \, , \label{Imf}
\end{eqnarray}
where $k$ is the incident momentum.  Note that as $q\to 0$, Eq. (\ref{Imf})
yields the optical theorem,
\begin{eqnarray}
    \sigma = \frac{4\pi}{k} {\rm Im} f(q=0)   \, .   \label{Opt}
\end{eqnarray}
At high energies, the imaginary part of the scattering amplitude can be
parametrized as
\begin{eqnarray}
   {\rm Im} f(q) = \frac{k\sigma}{4\pi} e^{\displaystyle{-q^2 r^2_\sigma/4}}
                 \, ;  \label{fq}
\end{eqnarray}
the slope parameter $r^2_\sigma$ depends weakly on energy, and at CERN
energies, $E_{lab}\simeq 200$ GeV, $r_\sigma\simeq 1.2$ fm \cite{MP}.
With this form we find
\begin{eqnarray}
    \frac{d\sigma}{d^2{b}}({\bf b})
     = \frac{\sigma}{\pi r^2_\sigma} e^{\displaystyle{-b^2/r^2_\sigma}}.
       \label{Pg}
\end{eqnarray}

    Substituting the two Gaussian parametrizations, (\ref{Cg}) and (\ref{Pg})
in the definition of $\alpha$, Eq.  (\ref{alfa}), and making use of the fact
that both $\xi$ and $r_s$ are $\ll R$, we find
\begin{eqnarray}
    \alpha = \frac{\rho_0\sigma}{(\pi r_\sigma^2)^2}
      \int dz \, e^{-z^2/2\xi^2}
      \int d^2 r_\perp d^2 r'_\perp
   \, e^{-({\bf r}_\perp-{\bf r}'_\perp)^2/2\xi^2}
      e^{-({r_\perp}^2 +{r'_\perp}^2)/r^2_\sigma},
\end{eqnarray}
which, by transforming to coordinates ${\bf r}_\perp\pm{\bf r}'_\perp$,
and using the relation (\ref{xir0}) is readily evaluated as
\begin{eqnarray}
    \alpha = \frac{1}{2\pi} \frac{\sigma}{\xi^2+r^2_\sigma} .\label{age}
\end{eqnarray}

    In order that the collision probability function, (\ref{Pg}), be less than
unity, the condition $\sigma\le\pi r^2_\sigma$ or equivalently $r_s\le
r_\sigma$ must be satisfied.  (Note that $r_s=(\sigma/\pi)^{1/2}$ determines
the strength of the interaction whereas $r_\sigma$ determines its range.)  For
the values for $r_\sigma$ and $\sigma$ quoted above $\sigma\le\pi r^2_\sigma$
is indeed satisfied.  This condition, used in (\ref{age}), implies that
$\alpha \le 1/2$.  This surprising result does not in fact disagree with the
earlier result $\alpha=1$ in (\ref{amax}) since there $d\sigma/d^2b$ was
assumed to be a step function.  Employing a diffuse form for $d\sigma/d^2b$
has the physical consequence that two nucleons lying on the same projectile
trajectory are not necessarily both hit, which softens the restrictions from
the correlations of the nucleons and results in a smaller value of $\alpha$.
With the actual values $\xi=0.75$ fm, $r_\sigma=1.2$ fm and $\sigma=32$ mb
given above, we find $\alpha=0.25$.

\subsection{Numerical studies}

    Pieper \cite{Pi} has calculated the detailed probability distribution
$P_N(b)$ for a proton colliding with various closed-shell nuclei numerically
using correlated many-body wave functions, with the assumption that
$d\sigma/d^2b$ is given by the step function of Eq.  (\ref{st}).  The
distribution function is considerably narrower for central collisions than a
Poisson.  By counting the number of nucleons in cylinders of radius 1 fm
(corresponding to $\sigma=32$ mb) he finds, for central events, $\alpha\simeq
0.31-0.33$ for both small (A=40) and large systems (A=224).  In addition, the
fluctuations are found to increase towards unity (corresponding to vanishing
$\alpha$) for peripheral events.  [Taking only the correlations of a free
Fermi gas of neutrons and protons into account yields $\alpha$ = 0.25 for
central collisions.] Pieper's value for $\alpha$ is larger than found above,
$\alpha=0.25$, primarily because of his use of a step-function form for
$d\sigma/d^2b$ instead of a more diffuse Gaussian.

\section{Fluctuations in nucleus-nucleus collisions}

    We now generalize the calcuation of the number of NN collisions and its
fluctuations to nucleus-nucleus collisions.  We continue to assume the Glauber
approximation with straight line geometry; the assumption that the nucleons do
not move between successive hits is valid at high energies as a consequence of
Lorentz contraction of the nuclei and the relatively small transverse momentum
transfers in NN collisions.

    The mean number of NN collisions in a central (${\bf b}=0$) collision of a
projectile nucleus B with a target nucleus A is given by
\begin{eqnarray}
    \langle N\rangle &=& \int d^3r d^3r'  \rho_B(r)\rho_A(r')
     \frac{d\sigma}{d^2{b}}({\bf r}_\perp-{\bf r}_\perp ') \nonumber\\
    & \simeq & \sigma \int dz \: dz' d^2s \rho_B({\bf s},z)\rho_A({\bf s},z')
         \nonumber\\
    &=& \sigma \int d^2s T_B(s)T_A(s),  \label{Nab}
\end{eqnarray}
where we assume $\sigma\ll {R_B}^2, {R_A}^2$.  When the projectile is
small, i.e., ${R_B}^2\ll {R_A}^2$, the mean number of collisions becomes
\begin{eqnarray}
    \langle N\rangle \simeq B\sigma\rho_0 R_A, \quad B\ll A \, ,
      \label{NBsmall}
\end{eqnarray}
which is just $B$ times the number of NN collisions in a central
proton-nucleus collision.

    The fluctuations in the number of NN collisions are found from
\begin{eqnarray}
    \langle N^2\rangle &=& \langle N\rangle  \nonumber\\
    &&+\, \int d^3r d^3r''d^3r'''\rho_B(r)\rho^{(2)}_A({\bf r}'',
   {\bf r}''')
     \frac{d\sigma}{d^2{b}}({\bf r}_\perp-{\bf r}_\perp '')
     \frac{d\sigma}{d^2{b}}({\bf r}_\perp-{\bf r}_\perp ''') \nonumber\\
    &&+\, \int d^3r d^3r'd^3r'' \rho^{(2)}_B({\bf r},{\bf r}')\rho_A(r'')
     \frac{d\sigma}{d^2{b}}({\bf r}_\perp-{\bf r}_\perp '')
     \frac{d\sigma}{d^2{b}}({\bf r}_\perp '-{\bf r}_\perp '') \nonumber\\
    &&+\, \int d^3rd^3r'd^3r''d^3r'''
          \rho^{(2)}_B({\bf r},{\bf r}')\rho^{(2)}_A({\bf r}'',{\bf r}''')
     \frac{d\sigma}{d^2{b}}({\bf r}_\perp-{\bf r}_\perp '')
     \frac{d\sigma}{d^2{b}}({\bf r}_\perp '-{\bf r}_\perp ''').
          \label{Nab2}
\end{eqnarray}
The four terms in Eq.  (\ref{Nab2}) correspond to:  1) the same projectile
nucleon colliding with the same target nucleon, 2) the same projectile at
position ${\bf r}$ colliding with two different nucleons at positions ${\bf
r}''$ and ${\bf r}'''$ in the target, 3) a nucleon at ${\bf r}''$ in the
target being hit by the two different nucleons at ${\bf r}$ and ${\bf r}'$ in
the projectile nucleus, and finally, 4) two different nucleons at positions
${\bf r}$ and ${\bf r}'$ in the projectile nucleus colliding with two
different nucleons at positions ${\bf r}''$ and ${\bf r}'''$ in the target
nucleus.

     The second term in (\ref{Nab2}) is evaluated analogously to the
proton-nucleus case and is
\begin{eqnarray}
    = \sigma\int d^2s T_B(s)T^2_A(s) \, -\langle N\rangle\alpha,
   \label{T2}
\end{eqnarray}
while the third term in (\ref{Nab2}) contains the two-body density
distribution of nucleus B and gives a contribution
\begin{eqnarray}
    = \sigma\int d^2s T^2_B(s)T_A(s) \,  -\langle N\rangle\beta,
   \label{T3}
\end{eqnarray}
where $\beta\simeq\alpha$ is the correlation parameter for nucleus B,
defined analogously to Eq.  (\ref{alfa}).

    The fourth term in Eq.  (\ref{Nab2}) reduces to
\begin{eqnarray}
   = \langle N\rangle^2 -\sigma\int d^2{\bf s} T_B(s)T^2_A(s)
              -\sigma\int d^2{\bf s} T^2_B(s)T_A(s)
              +\gamma \langle N\rangle \, .   \label{T4}
\end{eqnarray}
The four terms here arise from each nuclear two-body density distribution
having two terms [cf.  (\ref{rho2})]; they correspond respectively to the
product:  i) without any nuclear correlation functions, ii) with one
correlation function from nucleus A, iii) with one correlation function for
nucleus B, and finally iv) the product containing both correlation
functions.  The first three terms in (\ref{T4}), which contain only one or no
nuclear correlation function, are straightforward to calculate and do not
depend on the detailed forms of the correlation functions or the collision
probability function, as long as $\xi$ and $r_\sigma$ are much smaller than
the nuclear radii.  The last term is more tedious to evaluate because it
contains the products of both correlation functions and two collision
probabilities; these enter in the parameter, $\gamma$, defined by:
\begin{eqnarray}
   \gamma &=& \langle N\rangle^{-1} \int d^3rd^3r'd^3r''d^3r'''
         \, \rho_B(r)\rho_B(r')\rho_A(r'')\rho_A(r''') \nonumber\\
     &&  \times C_B({\bf r-r}') C_A({\bf r}''-{\bf r}''')
     \frac{d\sigma}{d^2{b}}({\bf r}_\perp-{\bf r}_\perp '')
     \frac{d\sigma}{d^2{b}}({\bf r}_\perp '-{\bf r}_\perp ''') \, .
\end{eqnarray}
With the Gaussian forms for the correlation function, Eq. (\ref{Cg}), and
for the scattering probability, Eq.  (\ref{Pg}), we find
\begin{eqnarray}
    \gamma = \frac{1}{2\pi} \frac{\sigma}{r^2_\sigma+2\xi^2}
           = \alpha \frac{r^2_\sigma+\xi^2}{r^2_\sigma+2\xi^2}.
            \label{gamma}
\end{eqnarray}

    The variance in the number of binary collisions, found from Eqs.
(\ref{Nab2})-(\ref{T4}), is
\begin{eqnarray}
    \omega_N= 1+\gamma-\alpha-\beta. \label{oAB}
\end{eqnarray}
This result generalizes the proton-nucleus result, Eq.  (\ref{o2}), to
nucleus-nucleus collisions, and improves the result derived in \cite{PRL91}
which assumed $\gamma=1$.  With $\xi=0.75$, $r_\sigma=1.2$ fm and $\sigma=32$
mb we find $\alpha=\beta=0.25$, $\gamma=0.15$, and $\omega_N=0.65$.

    The derivation of Eq.  (\ref{oAB}) assumes that the correlation length,
$\xi$, is much smaller than the nuclear radii, $R_B$ and $R_A$, as is the case
for a strongly correlated nuclear fluid.\footnote{ \baselineskip=15pt The
assumption $\xi\ll R_B$, is of course not valid for small projectile nuclei;
in the limit $B\to 1$, the term (\ref{T4}) vanishes, and (\ref{oAB}) reduces
to the result for proton-nucleus collisions, Eq.  (\ref{o2}).} The importance
of including the correct correlations is well-illustrated by calculating the
fluctuations in the opposite limit in which the nucleons are allowed to swarm
around in the nucleus as a free gas.  The correlation functions are then
$C_A=1/A$ and $C_B=1/B$, and when $r_\sigma\ll R_B \ll R_A$, one obtains
directly from (\ref{Nab2}) that
\begin{eqnarray}
    \langle N^2\rangle  & = &  \langle N\rangle
        + \sigma^2\int d^2 s T_B(s)T^2_A(s)
            \left(1-\frac{1}{A}\right)   \nonumber\\
     && + \sigma^2\int d^2 s T^2_B(s)T_A(s)
            \left(1-\frac{1}{B}\right)
        +  \langle N\rangle^2 \left(1-\frac{1}{A}\right)
           \left(1-\frac{1}{B}\right),
           \label{eq:n2gas}
\end{eqnarray}
where the terms correspond to the successive terms in (\ref{Nab2}).  The
resulting expression for the variance follows from Eqs. (\ref{Nab2})
and (\ref{eq:n2gas}):
\begin{eqnarray}
  \omega_N = 1 - \frac{B}{A} N_{pA} + \frac{B-1}{B} N_{pB}
        \quad {\rm (nuclear\,\, gas)}, \label{gasfluc}
\end{eqnarray}
where we define
\begin{eqnarray}
    N_{pA} &=& \langle N\rangle^{-1}\sigma^2\int d^2 s T_B(s)T^2_A(s),
                 \label{NpA} \\
    N_{pB} &=& \langle N\rangle^{-1}\sigma^2\int d^2 s T^2_B(s)T_A(s).
     \label{NpB}
\end{eqnarray}

    The quantities $N_{pA}$ and $N_{pB}$ are straightforward to calculate for
small projectiles, since then $T_A(s)\simeq 2\rho_0R_A$;
$N_{pA}$ reduces to the number of NN collisions in central proton-nucleus
collisions,
\begin{eqnarray}
    N_{pA} \simeq 2\sigma\rho_0R_A  \, ,\quad B\ll  A \, , \label{npA}
\end{eqnarray}
and
\begin{eqnarray}
    N_{pB} \simeq \frac{3}{2}\sigma\rho_0R_B \, ,\quad B\ll A
          \, . \label{npB}
\end{eqnarray}
The factor 3/2 instead of 2 follows from the impact parameter averaging
necessary for the projectile nucleus in the limit $B\ll A$.

    To calculate $N_{pA}$ and $N_{pB}$ more generally, we first assume uniform
density nuclei with sharp surfaces.  Then from (\ref{N}) we find,
\begin{eqnarray}
   \langle N\rangle &=& 4\pi\sigma\rho_0^2\int_0^{R^2_B} d\, b^2
            \left(R_B^2-b^2\right)^{1/2} \left(R_A^2-b^2\right)^{1/2}
              \nonumber \\
          &=& \frac{3}{4}B\sigma\rho_0R_A \left[
          1+x^2-\frac{1}{2x}(x^2-1)^2\ln\left(\frac{x+1}{x-1}\right) \right],
          \label{NAB}
\end{eqnarray}
with $x=R_A/R_B$.  From (\ref{NpA}) we have
 \begin{eqnarray}
 N_{pA}= \langle N\rangle^{-1}\sigma^2  \int d^2s T_B(s)T^2_A(s)
       &=& 8\langle N\rangle^{-1}\sigma^2 \pi\rho_0^3\int_0^{R^2_B} d\, b^2
            \left(R_B^2-b^2\right)^{1/2}  (R_A^2-b^2) \nonumber \\
           &=& 4\langle N\rangle^{-1}\sigma^2 B\rho_0^2R_A^2
             \left(1-\frac{2}{5}x^2\right),
   \label{NpAa}
\end{eqnarray}
and from (\ref{NpB}),
\begin{eqnarray}
 N_{pB} = \langle N\rangle^{-1}\sigma^2 \int d^2s T^2_B(s)T_A(s)
    &=& 8\langle N\rangle^{-1}\sigma^2 \pi\rho_0^3\int_0^{R_B^2} d\, b^2
            (R_B^2-b^2) \left(R_A^2-b^2\right)^{1/2} \nonumber \\
           &=& 4\langle N\rangle^{-1}\sigma^2 B\rho_0^2R_B^2\left(
              x^3-\frac{2}{5}(x^5-(x^2-1)^{5/2}) \right).
       \label{NpBa}
\end{eqnarray}
For small targets, the assumption $R_B \ll R_A$ used in
deriving (\ref{npA}) and (\ref{npB}) overestimates $N_{pA}$ and $N_{pB}$, as
well as the fluctuations.  For $B=A$, and ${\bf b}=0$, the above results give
\begin{eqnarray}
   N_{pA}=N_{pB}= \frac{8}{5}\sigma\rho_0 R_{A}  \, .  \label{AA}
\end{eqnarray}

    A diffuse nuclear surface decreases the average density, as well as
decreases $N_{pA}$ and $N_{pB}$ from the values calculated above.  A numerical
calculation with a Woods-Saxon density distribution
$\rho=\rho_0/(1+\exp{(r-R)/\delta})$, where $R$ is determined from the total
number of nucleons in the nucleus and the surface diffuseness is chosen as
$\delta=0.55$, yields a further reduction of $N_{pA}$ and $N_{pB}$ of order
20\% for $A$=28 to 10\% for heavy nuclei, compared to the above results for a
sharp surface.  The effects are largest for small nuclei, where the nuclear
surface is relatively large.

    The nuclear gas result (\ref{gasfluc}) is considerably larger than that of
the strongly correlated nuclear liquid, Eq.  (\ref{oAB}), as can be seen in
Fig. 3. The discrepancy can be traced back to the differences in the
correlation functions.  The derivation of (\ref{oAB}) assumed that $\xi\ll
R_B\ll R_A$, whereas in (\ref{gasfluc}) we took the correlation function
$C_A=1/A$, which implies that $\xi$ is of the size of the nuclear radius.  The
uncorrelated gas result for $\omega$ {\it cannot} be obtained from (\ref{oAB})
in the nucleus-nucleus case by choosing appropriate $\alpha$ and $\beta$,
e.g., as the value $\alpha = \langle N\rangle/A$ in the proton-nucleus case,
(\ref{agas}).  The larger $\omega_N$ in the ideal gas case arises mainly from
the second term in Eq.  (\ref{eq:n2gas}).  The extra fluctuations are caused
by ``geometric correlations," that is, if one of the projectile nucleons
collides with very few, or with numerous, target nucleons, then there is high
probability that the other projectile nucleons will do the same since the
projectile nucleons are nearby spatially.  Correlations in nuclei, however,
reduce the fluctuations in the number of proton-nucleus collisions and
effectively cancel the geometric correlations.

    This exercise clearly shows the necessity treating nuclear correlations
carefully in numerical simulations of relativistic heavy-ion collisions.  In
simulations the nucleons are often positioned successively in the nuclei by
Monte-Carlo methods with the restriction that they should be at least a
distance, $\tilde{\xi}$, apart.  Such a procedure, as discussed above,
describes a free gas with excluded volume, and fails to generate the
correlations characteristic of a self-bound liquid.  Generally, the
correlation function resulting from this procedure has a correlation hole of
radius $\tilde{\xi}$, but has a tail which oscillates for large distances and
can be negative.  While the correlation length, the range of the correlation
hole, is close to the interparticle spacing in nuclear matter, $\xi\simeq
r_0$, the minimum separation distance of the centers cannot be directly
translated into a correlation length.  When $ \tilde{\xi}>r_0$, the integral
of the density over the excluded volume exceeds unity, and the sum rule
(\ref{C}) requires the correlation function to be negative on average outside
the excluded volume.  (Note that the maximum value of $\tilde{\xi}$ is
$\raisebox{-.5ex}{$\stackrel{<}{\sim}$}$ 2$r_0$ in order to accomodate all the
nucleons in the nucleus).  When $\tilde{\xi}<r_0$ the correlation function has
a long tail characteristic of an interacting nuclear gas similar to the
``excluded volume" correlation function of Fig. 2.

    For example, the Fritiof model requires the centers to be at least 1.13 fm
apart \cite{Lund}.  Numerically, this results in $\alpha\simeq 0.4$, and so
for proton-nucleus collisions the fluctuations agree with those for the
more-correct correlation function.  However, in nucleus-nucleus collisions
this failure to include liquid-like correlations generally overestimates the
fluctuations, as Eq.  (\ref{gasfluc}) illustrates.  In the Fritiof model one
finds, for example, that $\omega_N\simeq 1.77$ in the case of $^{16}$O+Au
\cite{Lund}, a value between that of a free gas, Eq.  (\ref{gasfluc}), and
that for a nuclear liquid, Eq.  (\ref{oAB}).  The fact that a simulation
reproduces large fluctuations does not imply that it correctly includes the
physics producing the fluctuations.

    The density distribution resulting from positioning the nucleons by Monte
Carlo with a minimum separation can also have the unpleasant feature that it
is largest near the nuclear surface.  This is because the imposition of a
minimum distance between nucleons makes it harder to fit subsequent nucleons
into the nucleus; it is easier to accommodate the last ones near the surface
where there are fewer neighbors.

    A physically correct description of the initial nuclei should not only
produce the correct density distribution, but also describe nucleon-nucleon
correlations correctly, guaranteeing that the correlation function is
generally short ranged, vanishing within a distance less than the nuclear
radius, and does not become negative.  To implement correlations correctly in
numerically Monte-Carlo codes requires constructing the initial nuclei by
distributing all the particles in a nucleus at a time, rather than attempting
to simulate correlations according to a rule that puts the particles in
sequentially.  The correct probability distribution for the nucleon positions
is $|\Psi_A(r_1,r_2,...r_A)|^2$, where $\Psi_A$ is the $A$-particle ground
state nuclear wave function.  In practice, for given $|\Psi_A|^2$, one can
follow a standard Metropolis algorithm to generate most-probable
configurations.  Many-body wave functions of Jastrow form, represented as the
product of the mean-field wave function and two-nucleon correlators, readily
lend themselves to development of generators of initial conditions for
collision simulations.\footnote{
\baselineskip=15pt
In particular, S. Pieper (private communication) of Argonne National
Laboratory has generated useful simplified approximations to many-body nuclear
wave functions (\cite{Pi} and unpublished) of the Jastrow form that correctly
reproduce the A-particle spatial distribution functions, although not the
details of the spin correlations.  These wavefunctions have recently been
employed by Seki et al.  \cite{Seki} to study correlation effects in nuclear
transparency in (e,e'p) in heavy nuclei.}

\section{Color opacity and transparency, and cross-section fluctuations}

    In hadron-nucleus collisions at high energies, the internal configuration
of the color-carrying degrees of freedom of the projectile hadron are frozen
by Lorentz time dilation \cite{Mandelstam,Gribov}.  The hadron-nucleon cross
section is, as a consequence, characterized by a probability distribution
$P(\sigma)$.  The condition for a projectile nucleon to be frozen in passing
through a target nucleus is that the crossing time, $2R_A/c$, be less than the
internal dynamic time, given by the uncertainty relation between time and
energy \cite{FS},
\begin{eqnarray}
  2R_A \raisebox{-.5ex}{$\stackrel{<}{\sim}$}
  \frac{1}{\Delta E} \simeq \frac{2p_{lab}}{m_{N^*}^2-m_N^2}\simeq
   \frac{\gamma}{m_{N^*}-m_N}, \label{frozen}
\end{eqnarray}
where $m_N$ is the nucleon mass, N$^*$ is the lowest-lying resonance with
the quantum numbers of the nucleon, of mass $m_{N^*}\simeq$1.5 GeV, and
$\gamma$ is the Lorentz factor.  This condition is well-satisfied for
projectile energies above $p_{lab}\,\raisebox{-.5ex}{$\stackrel{>}{\sim}$}$ 40
GeV/c for heavy nuclei, as at the CERN SPS and LHC, and RHIC.  The concept of
cross-section fluctuations is well established in inelastic shadowing and
diffractive hadron-hadron scattering, as well as in coherent
diffractive-dissociation of hadrons scattering on nuclei, and is an essential
feature of the dynamics of ultrarelativistic heavy-ion
collisions \cite{PRL91,hadr}.

    The effects of cross-section fluctuations are several-fold.  When a hadron
is in a ``small-sized configuration" its interactions are suppressed because
of the small spatial extent of the color fields in the hadron, which leads to
the phenomenon of color transparency \cite{MB82}.  On the other hand, a hadron
in a frozen ``large hadronic configuration" experiences a stronger interaction
characterized by a cross section $\sigma > \bar{\sigma}$, its average value,
when passing through a nucleus, a ``color opacity" effect.  We now discuss, in
further detail than in Ref.  \cite{PRL91}, and also for the more general case
where the projectile B is not necessarily much smaller than the target A, how
these effects lead to enhancement of the fluctuations in observables such as
transverse energy and multiplicity.  We then compare with fluctuations found
at CERN.

    In calculating average quantities such as the mean number of collisions,
$\langle N\rangle$, the cross section $\sigma$ can be replaced by its average
value, $\bar{\sigma}=\langle\sigma\rangle_I\simeq 32$mb, where the subscript
$I$ refers to the average over internal configurations.  However, in
calculating fluctuations, one must first calculate for a given internal
configuration and then average the final result.  In particular the average of
the square of the cross section for collisions of the projectile p with
successive target nucleons $j$ and $j'$ is given by\footnote{
\baselineskip=15pt
We assume that the internal degrees of freedom decouple from the nuclear
wave function coordinates, i.e., we neglect small modifications of the
properties of nucleons in nuclei as revealed by the EMC effect.  We also do
not distinguish here the dispersion of the inelastic and total cross sections,
since they are very close.}
\begin{eqnarray}
\langle\sigma_{pj}\sigma_{pj'}\rangle_I\equiv\bar{\sigma}^2(1+\omega_\sigma)
\,,   \label{ss}
\end{eqnarray}
and the scaled variance of the cross-section fluctuations,
$\omega_\sigma$, is given by
\begin{eqnarray}
    \omega_\sigma\equiv  \frac{\langle\sigma^2\rangle_I}{\bar{\sigma}^2} -1.
           \label{oms}
\end{eqnarray}
The freezing of the projectile configuration during the passage across the
nucleus correlates $\sigma_{pj}$ and $\sigma_{pj'}$ through the internal
coordinates of the projectile.  When the projectile is in a small
configuration it has small probability to scatter on {\it both} $j$ and $j'$,
and when in a large configuration is has a large probability to do so.  [On
the other hand, a slow projectile has ample time to change its internal
structure between collisions; subsequent collisions are uncorrelated and then
$\langle\sigma_{pj}\sigma_{pj'}\rangle_I=\bar{\sigma}^2$, so that
$\omega_\sigma = 0$.] As we showed in Refs.
\cite{PRL91,QM91,hadr,BBFS93,Bor}, the size and form of the cross-section
fluctuations can be extracted from diffractive scattering experiments.

    Fluctuations of the color degrees of freedom of a hadron lead to
fluctuations in the overall scale as well as shape and range of its
interaction cross-section $d\sigma/d^2b$ with other hadrons.  In the following
we shall assume for simplicity that only the scale $\sigma$ of the
cross-section fluctuates but not its shape or range $r_\sigma$ [see, e.g., Eq.
(\ref{Pg})].  Fluctuations in shape and range can lead to further corrections
to the effects we found from including the finite interaction range, which
should be modelled in the future.  Employing the parametrization (\ref{Pg}),
we see that $\sigma^2$ enters as a common prefactor in (\ref{N2}); inclusion
of cross-section fluctuations at this level thus simply multiplies the right
side of (\ref{N2}) by a factor $1+\omega_\sigma$, and we directly obtain the
scaled variance in proton-nucleus collisions,
\begin{eqnarray}
  \omega_N=1-\alpha+\omega_\sigma(\langle N\rangle-\alpha);
   \label{os}
\end{eqnarray}
the important new ingredient is the last term due to the fluctuations in cross
sections.  For a central p-$^{208}$Pb collision, $\langle N\rangle \simeq 7$,
and with $\omega_\sigma\sim$ 0.2-0.3 at high energies as estimated in Ref.
\cite{hadr}, cross-section fluctuations contribute $\sim$ 1.5-2.0 to
$\omega_N$, thus increasing $\omega_N$ by a factor 2-3.

    The fluctuations in nuclear collisions can be calculated analogously to
the proton-nucleus case.  In the four terms of Eq.  (\ref{Nab2}), only the
second and third involve multiple collisions of the same nucleon; therefore in
these terms we must replace the prefactor
$\sigma^2$ by $\bar{\sigma}^2(1+\omega_\sigma)$.  The fluctuations in the
number of NN collisions in central nucleus-nucleus collisions for $\xi<R_B$
are thus given by
\begin{eqnarray}
    \omega_N
    =1+\gamma-\alpha-\beta+ \omega_\sigma(N_{pA}+N_{pB}-\alpha-\beta)
    \, . \label{osAB}
\end{eqnarray}
We can interpret the $\omega_\sigma$-term as the correlation coming from
each projectile nucleon making on average $N_{pA}-\alpha$ hits in the target,
and a target nucleon being hit on average $N_{pB}-\beta$ subsequent times by
projectile nucleons.  Equation (\ref{osAB}) generalizes the proton-nucleus
result of Eq.  (\ref{os}) to nuclear collisions by adding the correlations
from target nucleons being hit multiple times.  Expressions for $N_{pA}$ and
$N_{pB}$ are given in the previous section, Eqs.  (\ref{NpA})-(\ref{AA}).

\section{Transverse energy spectra}

    We show in this section the importance of correlations and cross-section
fluctuations by calculating their contribution to the fluctuations observed in
transverse energy spectra in ultrarelativistic heavy-ion collisions, e.g., by
NA34 \cite{NA34}.  Different microscopic models of nucleus-nucleus collisions
use either the number of binary NN-collisions or the number of participating
nucleons for the sources of particle or $E_T$ production \cite{PJD}.  We first
discuss the fluctuations in a binary collision picture, and then turn to the
participant, or ``wounded," nucleon picture.

\subsection{Fluctuations in Binary collisions}

    In an independent-source description of collisions, the $E_T$-distribution
is obtained by folding the $E_T$ contributions from $N$ sources, where the
distribution of the number of sources has mean $\langle N\rangle$.  The
overall form of the transverse energy distribution in a high-energy
nucleus-nucleus collision is determined by the geometry of the collision, as
is well-illustrated in the NA34 geometric model, where the $E_T$ distribution
is obtained by folding the $E_T$ contribution from $N$ sources (which becomes
Gaussianly distributed for sufficiently large $N$), multiplying by the
distribution of sources $d\sigma/dN$, and summing over $N$:
\begin{eqnarray}
   \frac{d\sigma}{dE_T} = \int d N \frac{d\sigma}{d N}
   \frac{e^{-(E_T-N\epsilon_0)^2/2\omega N\epsilon_0^2}} {(2\pi\omega
   N\epsilon_0^2)^{1/2}}.   \label{Et}
\end{eqnarray}
The parameter $\epsilon_0$ is interpreted in the NA34 model as the average
$E_T$ produced per source, while $\omega$ describes the standard deviation or
fluctuation around the mean value; both parameters are extracted by fitting to
data.  We do not take this interpretation literally here, but rather adopt the
point of view that the model provides a useful parametrization of the data.

    The cross section for making $N$ sources, which is taken to be the number
of binary collisions, is
\begin{eqnarray}
    \frac{d\sigma}{dN} = \int d^2 b P_N(b),
\end{eqnarray}
where $P_N(b)$ is the probability for making $N$ sources in a collision at
impact parameter $b$.  In the simplest geometric model, $P_N$ is non-zero only
at the mean number of sources, $\bar{N}(b)$, for given impact parameter; for
example, for small projectiles $B$ and spherical target nuclei of radius
$R_A$, one has $\bar{N}(b)=2B\bar{\sigma}\rho_0\left(R_A^2-b^2\right)^{1/2}$.
More generally one can take $P_N(b)$ to be a Poisson distribution around the
mean, but, as we have discussed above, correlations lead in fact to a much
narrower structure for $P_N$ than Poisson.  Equation (\ref{Et}) also describes
the multiplicity distribution when $E_T$ is replaced by the multiplicity $n$;
the parameter $\epsilon_0$ is then replaced by the average number of particles
produced per source, and $\omega$ becomes the corresponding scaled variance.

    Our calculations above for the fluctuations in collision numbers are for
given impact parameter $b$; averaging over impact parameters broadens the
spectra and increases the fluctuations.  The effect of impact parameter
averaging is in fact separated out in the NA34 analyses of nucleus-nucleus
collisions \cite{NA34}.  As described in Ref.  \cite{BFH}, the tail of the
$E_T$ distribution in ultrarelativistic heavy-ion collisions is determined by
central events, on which we concentrate here.\footnote{ \baselineskip=15pt The
procedure of extracting fluctuations at given impact parameter is much more
difficult for proton-nucleus collisions which do not exhibit the plateau
spectrum in nucleus-nucleus collisions.  While the characteristic ``shoulder"
in nucleus-nucleus $E_T$-spectra marks the beginning of region dominated by
central collisions, in proton-nucleus collisions impact-parameter averaging
results in a long tail which is further smeared by fluctuations as well as
degradation and rescattering \cite{HH}.} The NA34 analysis thus provides a
reliable, albeit model-dependent, way to extract the fluctuations $\omega$ for
central ($b=0$) collisions from data, and yields values of $\omega$ in
collisions of $^{32}$S on nuclear targets  which range, in the
pseudorapidity interval
$-0.1<\eta <2.9$,
 from $\sim$ 1.5 to $\sim$ 5.8.  The $\omega$'s extracted
from
the NA34 analyses directly describe the transverse-energy fluctuations in
central collisions,
\begin{eqnarray}
\omega \equiv \langle N\rangle
  \frac{\langle E_T^2\rangle-\langle E_T\rangle^2}{\langle E_T\rangle^2},
  \label{omega}
\end{eqnarray}
where $\langle N\rangle$ is evaluated at zero impact parameter.  The
problem is to understand the physics producing the large values of $\omega$
needed by NA34.  As we first discussed in Ref.  \cite{PRL91}, and expand on
here, cross-section fluctuations lead to a substantial contribution to
$\omega$, capable of explaining deviations of standard estimates from
experiment \cite{NA34,BFS}.

    The principal contributions to $\omega$ are:
\begin{eqnarray}
 \omega = \omega_0 +  \omega_{def}+ \omega_N,
  \label{omega2}
\end{eqnarray}
where $\omega_0$ is the width of the single source $E_T$-distribution,
$\omega_{def}=4\langle N\rangle\delta^2/45$ is the variance of the
fluctuations from the deformation of the target, with $\delta$ the nuclear
deformation parameter, particularly large for W and U. The third term,
$\omega_N=(\langle N^2\rangle-\langle N\rangle^2)/\langle N\rangle$, takes
into account fluctuations in the number of sources in a central collision.
With the assumption that the number of sources is given by the number of
binary collisions, fluctuations in the number of sources is given by Eq.
(\ref{oAB}) in the absence of cross-section fluctuations, and Eq.
(\ref{osAB}) with cross-section fluctuations included.\footnote{
\baselineskip=15pt The experimentally-extracted values of $\omega$ include a
small contribution, $\omega_{res}$, from the finite energy-resolution of the
detectors, $\Delta E/\langle E\rangle \sim 0.45/\sqrt{E_{GeV}}$ (where
$E_{GeV}$ is the energy in units of GeV), which leads to $\omega_{res} =
(0.45)^2 \langle N\rangle/E_{GeV}$ of order 0.2; we are grateful to Dr.  H.
Str\"obele for emphasizing the existence of this contribution.  In Eq.
(\ref{omega2}) we also have neglected terms due to decrease of the
transverse-energy production per source in successive hits.  As argued in Ref.
\cite{BFS} these terms lead to a small reduction of $\omega$.}

    Using $\omega_0 \approx 0.5$ as determined from pp-collisions in the
comparable energy and rapidity range, one cannot, with these assumptions in
the absence of cross-section fluctuations, explain the experimentally observed
large fluctuations.\footnote{
\baselineskip=15pt
Rescattering of secondaries might be expected to broaden the tails of the
$E_T$-distribution and thus increase $\omega$, since it allows a secondary to
lead to a wider range of transverse energy and multiplicity in final states.
But rescattering also increases the number of scatterings and thus tends to
narrow the fluctuations; in general, adding more degrees of freedom to a system
while keeping its average output fixed decreases its fluctuations.  A
calculation for central collisions of $^{32}$S on Au at CERN-SPS energies by
Werner
\cite{Wer90} using the VENUS 4.16 code both with and without rescattering shows
no significant change in the relative width of the distribution.  We are
indebted to Dr.  K. Werner for providing us with the results of this
calculation.}
With $\bar{\sigma}\simeq32$ mb and $\rho_0\simeq0.15$ fm$^{-3}$, and a
heavy target we have $N_{pA}\simeq 2\bar{\sigma}\rho_0R_A\simeq 1.2 A^{1/3}$.
In CERN and Brookhaven experiments, with $^{16}$O or $^{32}$S projectiles,
$N_{pB}\simeq \frac{3}{2}\bar{\sigma}\rho_0R_B\simeq 2-3$.  We show in Fig. 4
$\omega - \omega_{def}=\omega_N+\omega_0$ given by Eq.  (\ref{osAB}), for
$^{32}$S, with various values of $\omega_\sigma$, taking $\omega_0=0.5$,
$\alpha=\beta=0.25$ and $\gamma=0.15$, as discussed above.  We see that
cross-section fluctuations are able to account for the large $\omega$'s found
experimentally with $\omega_\sigma\sim 0.25$, a value consistent with that
extracted from inelastic shadowing in nucleon-deuteron scattering as well as
forward diffractive scattering amplitudes at CERN-SPS energies \cite{hadr}.
Equation (\ref{osAB}) describes the $^{16}$O data \cite{Aa} as well with
similar values for $\omega_\sigma$.  To disentangle the color transparency and
opacity effects from other sources of $E_T$ fluctuations in future
experiments, especially rescattering, it will be useful to study the energy
and rapidity dependence of the $E_T$ fluctuations.

\subsection{Fluctuations in Participants or ``Wounded nucleons"}

    Let us consider briefly the fluctuations in a model in which the number of
sources is the number of participant or wounded nucleons.  The number of
sources in proton-nucleus collisions is very similar for the binary-collision
and the wounded-nucleon pictures, since the number of participants is just the
number of binary collisions plus one, the projectile.  Similarly, the
fluctuation around the mean number of sources is almost the same.  For nuclear
collisions the numbers of sources in the two pictures can differ
considerably \cite{Evert}.

    The geometric-model result (\ref{Et}) can also be applied in the
wounded-nucleon picture.  The interpretation of the source parameters
$\epsilon_0$ and $\omega_0$ is, however, very different in the two pictures.
In the binary collision model $\epsilon_0$ is the transverse energy produced
in a nucleon-nucleon collision whereas in the wounded nucleon model it is half
that value.  As we see from Eq.  (\ref{omega}), $\Omega\equiv\omega/\langle
N\rangle$ is the same in both cases.

   The number of participants in a nucleus-nucleus collision is given by
\begin{eqnarray}
  \langle N\rangle_{WN} \simeq \int d^2 s
  \left[T_B({\bf s})(1-e^{-\sigma T_A({\bf s}+{\bf b})})
  +T_A({\bf s}+{\bf b})(1-e^{-\sigma T_B({\bf s})})\right],
     \label{WNM}
\end{eqnarray}
a formula which assumes $\sigma\ll\pi R_B^2$.  When the projectile is much
smaller than the target nucleus this result simplifies to
\begin{eqnarray}
   \langle N\rangle_{WN} \simeq B(1-e^{-N_{pA}}) +
     \pi R_B^2 T_A(b)\left(1-\frac{2}{\tilde{N}_{pB}^2}
    [1-(1+\tilde{N}_{pB})e^{-\tilde{N}_{pB}}]\right) \, ,
\end{eqnarray}
where $N_{pA}=\sigma T_A(b)$ and $\tilde{N}_{pB}=\sigma T_B(0)$.  While
$N_{pA}$ is the same as in Eq.  (\ref{NpA}), $\tilde{N}_{pB}$ is a factor 3/4
smaller than $N_{pB}$ in Eq.  (\ref{NpB}).

    In the limit in which all nucleons in the overlapping volume of the two
nuclei in the collision participate, i.e., small nucleon mean free-paths,
corresponding to large $N_{pA}$ and $\tilde{N}_{pB}$, and assuming sharp
nuclear surfaces we find from (\ref{WNM}) that for central collisions
\begin{eqnarray}
   \langle N\rangle_{WN} = B+ A-|A^{2/3}-B^{2/3}|^{3/2},
       \label{NWNBA}
\end{eqnarray}
For small projectiles (\ref{NWNBA}) reduces to
\begin{eqnarray}
  \langle N\rangle_{WN} \simeq B+\pi R_B^2 T_A(b) \, ,\quad B\ll A \, ,
\end{eqnarray}
valid in this form also for non-central impact parameters.  In comparison,
the mean number of binary collisions, $\langle N\rangle\simeq B\sigma T_A(b)$,
is generally is larger.

    The latter results are, to first approximation, independent of the cross
section; therefore we do not expect fluctuations in $\sigma_{ij}$ in this
limit to affect the distribution of wounded nucleons significantly.

    To estimate the fluctuations in the number of participants including
correlations, we limit ourselves for simplicity to central collisions and
assume sharp nuclear surfaces and the limit of short nucleon mean free paths.
In this limit, with B $\le$ A, all nucleons from nucleus $B$ participate and
so there are no fluctuations in that number.  The fluctuation in the number of
participants in nucleus $A$ is determined from
\begin{eqnarray}
    \langle N(N-1)\rangle_{WN\in A} = \int d^3r d^3r' \, \rho_A^{(2)}
    ({\bf r,r}')
    \theta(R_B-|{\bf r}_\perp|)\theta(R_B-|{\bf r}'_\perp|)
    \, . \label{N2WN}
\end{eqnarray}
[Note the formal similarity of this equation to Eq.  (\ref{N2a}) for pA
collisions with the range $r_s$ of the nucleon-nucleon interaction replaced by
$R_B$.] For an uncorrelated nuclear gas, for which $\xi=R_A$ and $C_2=1/A$,
Eq.  (\ref{N2WN}) gives $=\langle N\rangle_{WN\in A}^2(1-1/A)$ and so
$\omega_N=1-\langle N\rangle/A$, the usual binomial result (cf. \ref{agas})).
More realistically, $\xi\ll R_B$, in which case Eq.  (\ref{N2WN}) implies
$\langle N^2\rangle _{WN\in A} =\langle N\rangle_{WN\in A}^2$ and thus
$\omega_N=0$ (cf. (\ref{amax})).  Correlations effectively eliminate
fluctuations in $N_{WN}$.  Fluctuations in observables arise only from
fluctuations in the individual sources, given by $\omega_{0,WN}=2\omega_0$,
since two wounded nucleons make up one binary NN collision.  To compare the
predictions of fluctuations in the wounded-nucleon picture to the NA34 data,
we should keep in mind the conventional but arbitrary scaling by $\langle
N\rangle$ in the relation of (\ref{omega}) to the fluctuations in transverse
energy, and thus compare $\omega_{WN}=2\omega_{0,WN}\langle N\rangle/\langle
N\rangle_{WN}\simeq 1-2$ to the $\omega$ extracted in the NA34 geometric
model.  The fluctuations in the wounded-nucleon picture are generally not
large enough to explain the NA34 data.

\section{Conclusion}

    As we have seen, nucleon correlations in projectile and target nuclei, the
range of the underlying NN interactions, and cross-section fluctuations are
all important for determining fluctuations in the number of NN collisions and
thus observables in nuclear collisions.  Correlations reduce the density
fluctuations and therefore also the fluctuations in the number of NN
collisions.  Because the nucleus is a self-bound liquid of nucleons,
correlations are particularly important in nucleus-nucleus collisions, and
lead to significant reduction in fluctuations compared to those for a gas of
nucleons assumed freely swarming in the nucleus.  Interestingly, the
correlations in the incident nuclei are one of the few features of the
original nuclear structure that are a determining factor in high-energy
collisions.  Allowing for a larger range for the interaction or collision
probability reduces the effect of the correlations between two nearby
nucleons, and increases the fluctuations.  Cross-section fluctuations are an
important feature of collisions of a hadronic projectile at sufficiently high
energy that time dilation does not allow the projectile to change its
configuration between successive collisions in the target nucleus.  In this
case successive collisions become correlated, with the effect of increasing
the fluctuations in the number of NN collisions significantly.

    We emphasize the need to incorporate the physics we discussed here into
Monte-Carlo event generators for AA collisions.  The method currently applied
in many Monte-Carlo codes of constructing initial conditions by successively
filling a nucleus with nucleons separated by a minimum distance does not
reproduce nuclear correlations and fluctuations satisfactorily, especially in
nucleus-nucleus collisions; approaches to implementing nuclear correlations
that correctly describe a nuclear liquid were discussed at the end of Sec. 3.
To a first approximation, cross-section fluctuations can be incorporated in
the probabilistic language of Sec. 3 by assuming factorization for
interactions of two nucleons (1 and 2) in different configurations, i.e.,
assuming the cross section to have the form,
$\sigma=\sigma_1\sigma_2/\bar{\sigma}$ (cf. discussion in \cite{PRL91,QM91}),
and then generating $\sigma$ for each of the nucleons with individual weights
$P(\sigma_1)$ and $P(\sigma_2)$, where $P(\sigma)$ is the normalized
distribution of cross sections for NN scattering \cite{hadr}.

    Beyond fluctuations in the probability of collisions, one should also
incorporate fluctuations in the outcomes of individual collisions, i.e.,
fluctuations on the parton level.  Since parton distributions depend on the
size of the nucleon configuration, fluctuations in internal nucleon
configurations lead directly to fluctuations in parton distributions and hard
parton-parton interactions.  While microscopic collision models in terms of
partons have been developed for ultrarelativistic heavy-ion collisions (see,
e.g., \cite{KKG,EKL}), they so far only take fluctuations in number of
collision partners into account, employing average structure functions, or
parton momentum distribution functions, for individual nucleons.  A simple way
to take fluctuations on the parton level in account in such models, based on
the fact that increase of the effective transverse size of a hadron allows
emission of softer partons, is to let the parton distributions
$p_N(x,Q^2,\sigma)$ scale as
\begin{eqnarray}
 p_N(x,Q^2,\sigma) = p_N(x,Q^2 \sigma/\bar{\sigma}),
\label{close}
\end{eqnarray}
similar to the rescaling model of the EMC effect \cite{Close}.

    At small $x$ one also has to take into account shadowing and enhancement
effects which depend both on $\sigma$ and the impact parameter of the
colliding nucleon \cite{FLS90}; discussion of these effects is beyond the
scope of this paper.  Furthermore, observed fluctuations in hard parton-parton
interactions will depend on the subset of events on which one triggers in a
collision (see \cite{BF93} and discussion in \cite{Bor}), e.g., biasing
towards high $E_T$ events will enhance the probability of large interaction
cross-section configurations occurring in the colliding nuclei.  Methods to
study this phenomenon experimentally in pA collisions are discussed in
\cite{FS85}.

\acknowledgments
    This work was supported by NSF Grants PHY 89-21025 and PHY 94-21309, DOE
Contract DE-AC03-76SF00098, DOE Grant DE-FG02-93ER40771, and Binational
Scientific Foundation (BSF) Grant 9200126.  BB is grateful for support from
the Alexander von Humboldt Stiftung and HH from the Danish Natural Science
Research Council.  We thank Steven Pieper, Vijay Pandharipande, Heinz Sorge,
and Evert Stenlund for useful discussions.

\vspace{1cm}
\appendix{The AGK cutting rules and binary collisions}

    The AGK diagram cutting rules for high-energy scattering processes
\cite{AGK,BT} lead to the theorem that in nucleon-nucleus (NA) scattering,
inclusive particle production away from the projectile fragmentation region is
simply that in NN scattering times the number of nucleons, $A$, in the target
nucleus.  In this Appendix we briefly discuss the extension of this result to
nucleus-nucleus collisions at central rapidities, to give an indication of how
the probabilistic methods we have used in this paper can be extracted from a
more formal field-theoretic approach.  The prediction of the theorem for
nucleus-nucleus collisions that particle production scales with the number of
NN collisions supports the use in this paper of the binary collision model in
describing fluctuations in the central rapidity range.

    The theorem is based on the Glauber multiple NN scattering formalism
within the eikonal approximation, and thus ignores rescattering of produced
particles during the collision.  Rescattering of target nucleons on other
target nucleons is not taken into account, and in nucleus-nucleus collisions
neither is scattering of projectile nucleons on other projectile nucleons.  In
addition, coherent processes, such as showers of particles overlapping
spatially and interacting strongly, are ignored, as is a possible transition
to a quark-gluon plasma in which particle production cannot be described in
terms of nucleon-nucleon interactions.  Finally, the AGK theorem ignores
constraints from global energy conservation in multiple NN collisions, and is
thus valid only at those rapidities and at sufficiently large energies that
particle production depends slowly on the initial energy.  Within these
assumptions, AGK show that there is an exact cancellation of processes
involving inelastic interactions of the projectile with more than one nucleon
at a time.  For example, in the interaction of the projectile with two
nucleons, the forward amplitudes for the two processes where the projectile
interacts inelastically with one of the nucleons and elastically with the
other cancels against the amplitude for production of particles in inelastic
interactions with both nucleons.  The net result is that the total
multiplicity in inelastic collisions is the sum of the multiplicities in
inelastic collisions of the projectile with individual nucleons.

    In the projectile fragmentation region in NA scattering, global energy
conservation constrains particle production; the cancellation fails in the
projectile fragmentation region because the way in which the projectile energy
is shared in inelastic collisions depends on the number of target nucleons
with which it collides.  The net result is shadowing of particle production
close to the projectile rapidity.\footnote{\baselineskip=15pt One can, in
fact, extend the AGK arguments to include energy conservation within the
Glauber model, and derive descriptions of the bulk characteristics of
hadron-nucleus scattering in good agreement with available data ($E_{inc} \le
400$ GeV) [reviewed in \cite{Sha}].} The more nucleons involved in a
collision, the stronger is the constraint of global energy conservation, and
the wider the rapidity interval where the AGK cancellation is not effective.
For example, in central nucleus-nucleus collisions at lab energy of 200 GeV/A,
as at CERN, the limited total energy constrains particle production in the
nuclear fragmentation regions, and precludes application of the AGK theorem
there.

    The extension of the AGK theorem to nucleus-nucleus (BA) collisions, for
particle production away from nuclear fragmentation regions, is:
\begin{equation}
  \sigma^{BA\to a+X} = BA \sigma^{NN\to a+X}  \, , \label{sAGK}
\end{equation}
where $\sigma^{BA\to a+X}$ and $\sigma^{NN\to a+X}$ are the inclusive
cross sections for producing particle {\it a} in BA and NN collisions
respectively.  The number of particles {\it a} produced per collision is
\begin{equation}
  \nu^{BA\to a} = \frac{\sigma^{BA\to a+X}}{\sigma^{BA}}  \, , \label{nuAGK}
\end{equation}
where $\sigma^{BA}$ is the total production cross section for colliding
nuclei A and B (i.e., not including quasielastic processes in which no hadrons
are produced); for NN scattering $\sigma^{NN}=\bar{\sigma}$.  The mean number
of NN collisions averaged over impact parameter is then
\begin{equation}
  \langle N\rangle = \frac{\nu^{BA\to a}}{\nu^{NN\to a}}
     = BA\frac{\sigma^{NN}}{\sigma^{BA}},
    \label{NAGK}
\end{equation}
a result in fact independent of the particle $a$.  By writing
this equation as
\begin{equation}
  \langle N\rangle
     = \int \frac{d^2b}{\sigma^{BA}} \quad \sigma^{NN}\int d^2s
       T_A({\bf s}+{\bf b})T_B(s),
       \label{NbAGK}
\end{equation}
we may interpret Eq. (\ref{NAGK}) as an average of impact parameter of
the number of collisions at given impact parameter,
\begin{equation}
  \langle N(b)\rangle
= \sigma^{NN}\int d^2s T_A({\bf s}+{\bf b})T_B(s);
\label{NbAGK1}
\end{equation}
cf.  Eq.  (\ref{Nab}).
The generalization of Eq.  (\ref{NbAGK}) to a finite interaction range is
outlined and applied to diffractive scattering and inelastic shadowing effects
in Refs.  \cite{Gribov67,QM91,hadr}.

    The above example indicates the correspondence between the probabilistic
language of Sec. 3 and field theory.  The AGK theorem applied to
nucleus-nucleus collisions leads to the same partial probabilities as the
binary collision model of collisions involving a given number of nucleons in
inelastic interactions.  Binary collision models make particular assumptions
about how energy is shared in the various subcollisions.  At central
rapidities, where the energy constraints are unimportant, the results are
independent of these assumptions, and one can apply the AGK theorem.  However,
binary collision models give specific predictions in the fragmentation regions
as well.

    At very high energies the contribution of minijets to particle and $E_T$
production starts to become comparable to that from soft processes.  According
to Eskola et al.  \cite{EKL} minijets contribute more than 50\% at energies
above RHIC energies, $\sqrt{s} \raisebox{-.5ex}{$\stackrel{>}{\sim}$} 200$GeV
(such estimates, however, carry the uncertainty of how low in transverse
momentum one can use perturbative qcd to calculate jet production).  Since
minijets are hard processes with kinematics where nuclear shadowing in the
parton distributions can be neglected, they are produced independently in each
NN collision.  Hence in this case the binary collision model for parton-parton
scattering should be valid, although one has to take into account fluctuations
in the parton distributions on nucleons, cf.  Eq.  (\ref{close}).

    The inclusion of color fluctuation effects in the description of
nucleus-nucleus collisions for calculation of total cross sections is
equivalent in qcd to inclusion of Gribov inelastic shadowing effects, the
presence of inelastic intermediate states in the inelastic eikonal
approximation (at the level where one neglects the high-mass diffraction,
described by {\it triple Pomeron} diagrams).  The use of the eigenstate
formalism \cite{FP56,GW60} (applied in this context in Refs.  \cite{hadr,Bor})
allows one to diagonalize the transition matrices in the inelastic eikonal
diagrams and write these diagrams as a sum of terms corresponding to the
scattering of nucleons in the states with fixed $\sigma$.  The probabilistic
interpretation of the eikonal approximation for the multiple scattering
combinatorics is thus preserved even in the presence of color fluctuations,
allowing one to understand the results of Sec. 3 from a field-theoretic
perspective.  For example, Eq.  (\ref{Nab2}), with color fluctuations, can be
derived using the AGK rules for the double inclusive spectrum.  Indeed, any
quantity calculated in the eikonal approximation proportional to $\sigma^n$ is
given, with color fluctuations taken into account, by the same expression with
the substitution $ \sigma^n \rightarrow \langle\sigma^n\rangle$.

    We also note that the AGK cutting rules connect the cross sections for
inelastic incoherent hadron production and diffractive hadron production, as
arising from different cuts of the same diagrams.  Hence coherent hadron
diffraction in hadron-nucleus collisions can be calculated using the same
distribution of cross sections $P(\sigma)$ \cite{FMS93} as used for inelastic
incoherent hadron production.  The agreement of the results of the calculation
of inelastic coherent diffractive processes in hadron-nucleus scattering with
existing data \cite{DAT} provides a further confirmation of the picture of
color fluctuations adopted here.


\figure{ \label{Glauber}
    Straight-line geometry of a proton-nucleus collision displaying the length
scales of the cross-sectional distance, $r_s$, the interparticle spacing,
$r_0$, and the correlation length, $\xi$.}

\figure{\label{Corfun}
    Correlation functions from \cite{Vijay} (full curve), the Gaussian fit,
Eq.  (\ref{Cg}) (dashed curve), an ideal gas, $C_A=1/A$ (dotted curve) and a
gas with excluded volume (long-dashed curve) for a nucleus with $A=208$.}

\figure{\label{ONs}
    Fluctuations $\omega_N$ for central collisions of $^{32}$S on different
targets for a strongly correlated nucleus, Eq.  (\ref{osAB}) (solid line), as
well as for a nuclear gas, Eq.  (\ref{gasfluc}) (dashed line).}

\figure{\label{Osigmas}
    Fluctuations $\omega - \omega_{def}=\omega_0+\omega_N$
for central collisions of $^{32}$S on
different targets for $\omega_\sigma= 0.0$, $0.1$, $0.2$ and $0.3$ calculated
with Eq.  (\ref{osAB}).  The experimental values (dots) are taken from Ref.
\cite{NA34}.  }

\end{document}